# The ASIMOV Prize for scientific publishing: HEP researchers trigger young people toward science

**W.M. Alberico** [a] , **R. Antolini** [b], **S. Arezzini** [c], **L. Bellagamba** [d], **N. Cavallo** [e], **C. Cecchi**[f], **S. Cherubini** [g], **R. Colalillo** [h], **G. Di Sciascio** [i], **C. Distefano**[j], **S. Fuso**[k], **G. Galati** [l], **R. Hueting** [m], **S. Leone** [c], **M. Lissia** [n], **S. Miozzi** [i], **D. Mura** [n], **A. Papa** [o], **A. Parisi**[p], **G.M. Piacentino** [q], **C. Puggioni** [n], **M. Radici** [r], **S. Sebastiani** [b], **A. Sidoti** [d], **L. Silvestris** [s], **M. Tuveri** [t], **F. Ursini** [b], **A. Ventura** [u,*], **E. Vigezzi** [v], **F. Vissani** [b], **D, Vitali** [w]

a    Istituto Nazionale di Fisica Nucleare, Sezione di Torino, Italy

b    Istituto Nazionale di Fisica Nucleare, Laboratori Nazionali del Gran Sasso, Italy

c    Istituto Nazionale di Fisica Nucleare, Sezione di Pisa, Italy

d    Istituto Nazionale di Fisica Nucleare, Sezione di Bologna, Italy

e    Università degli Studi della Basilicata & Istituto Nazionale di Fisica Nucleare, Sezione di Napoli, Italy

f    Università degli Studi di Perugia & Istituto Nazionale di Fisica Nucleare, Sezione di Perugia, Italy

g    Università degli Studi di Catania & Istituto Nazionale di Fisica Nucleare, Sezione di Catania, Italy

h    Università degli Studi di Napoli Federico II & Istituto Nazionale di Fisica Nucleare, Sezione di Napoli, Italy

i    Istituto Nazionale di Fisica Nucleare, Sezione di Roma Tor Vergata, Italy

j    Istituto Nazionale di Fisica Nucleare, Laboratori Nazionali del Sud, Italy

k    CICAP - Comitato Italiano per il Controllo delle Affermazioni sulle Pseudoscienze, Italy

l    Università degli Studi di Bari & Istituto Nazionale di Fisica Nucleare, Sezione di Bari, Italy

m    Deep Blue R&D, Rome, Italy

n    Istituto Nazionale di Fisica Nucleare, Sezione di Cagliari, Italy

o    Università della Calabria & Istituto Nazionale di Fisica Nucleare, Gruppo collegato di Cosenza, Italy

p    ALI - Associazione Librai Italiani, Italy

q    Università Telematica Internazionale UniNettuno, Italy

r    Istituto Nazionale di Fisica Nucleare, Sezione di Pavia, Italy

*Speaker

*s   Istituto Nazionale di Fisica Nucleare, Sezione di Bari, Italy*

*t   Università degli Studi di Cagliari, & Istituto Nazionale di Fisica Nucleare, Sezione di Cagliari, Italy*

*u    Università del Salento & Istituto Nazionale di Fisica Nucleare, Sezione di Lecce, Italy*

*v    Istituto Nazionale di Fisica Nucleare, Sezione di Milano, Italy*

*w   Università di Camerino & Istituto Nazionale di Fisica Nucleare, Sezione di Lecce, Italy*

*E-mail:* francesco.vissani@lngs.infn.it, andrea.ventura@le.infn.it

This work presents the ASIMOV Prize for scientific publishing, which was launched in Italy in 2016. The prize aims to bring the young generations closer to scientific culture, through the critical reading of popular science books. The books are selected by a committee that includes scientists, professors, Ph.D. and Ph.D. students, writers, journalists and friends of culture, and most importantly, over 800 school teachers. Students are actively involved in the prize, according to the best practices of public engagement: they read, review the books and vote for them, choosing the winner. The experience is quite successful: 12,000 students from 270 schools all over Italy participated in the last edition.

The possibility of replicating this experience in other countries is indicated, as was done in Brazil in 2020 with more than encouraging results.

## 1. Introduction

Isaac Asimov (1920–1992) was a well-known US writer, a professional scientist, a committed intellectual, an inspired promoter of culture and a great lover of knowledge [1]. He demonstrated that there is no need to create barriers between scientific and humanistic cultures. Starting from this principle, the ASIMOV Prize was created to celebrate this immortal example of dissemination of scientific culture.

## 2. A prize promoting scientific culture

The ASIMOV Prize was founded in Italy in 2016 with cultural and non-profit purposes, aiming to bring the young generations closer to scientific culture, through the critical reading of popular science books. Among the organizing bodies, the INFN (Italian Institute of Nuclear Physics) has played a central role since the the first edition of the prize, together with a growing number of universities and cultural and research institutions, under the guidance and coordination of the spokesperson, Prof. Francesco Vissani [2].

According to the ASIMOV Prize's rules, which are updated and consolidated over the years, in order to be considered, books are intended to be suitable for non-expert audiences and have to be published in Italian (or in translation) for the first time in the two years prior to the current edition. A National Committee has to select a shortlist of finalist books from a large list of proposed works. This committee is composed of scientists (coming from many different scientific communities), professors, Ph.D. and Ph.D. students, writers, journalists, friends of culture in general and, most of all, a large number of school teachers (about 800 as of 2022). Many scientific subjects are considered of interest for the ASIMOV Prize: they include chemistry, medicine, physics, neurophysiology, mathematics, computer science, astronomy, but also biology, geology, geography, engineering, anthropology, history and philosophy of science.

The National Committee of the ASIMOV Prize is the union of Regional Committees, working in synergy from all over Italy: Abruzzo, Basilicata, Calabria, Campania, Emilia-Romagna, Lazio, Liguria, Lombardy, Marche, Molise, Piedmont, Apulia, Sardinia, Sicily, Tuscany and Umbria. The participation in the Committee is on a voluntary and unpaid basis. Each member has the right to propose a list of books for the final selection, which is announced on the official website (https://www.premio-asimov.it/), typically around the start of the school year.

The Regional Committees contact the High Schools with which they collaborate and communicate the finalist books for the current edition of the ASIMOV Prize. The national jury is composed of students from Italian high schools participating in the prize. Each student-juror has to register on the website, choose, read, review and vote (rating from 1 to 10) one or more of the finalist books. Since the first edition in 2016, the number of High Schools in Italy participating in the prize has been constantly growing: presently there are 272 schools on the entire national territory (distributed as shown in Fig. 1), for a total of 12,326 students-jurors.

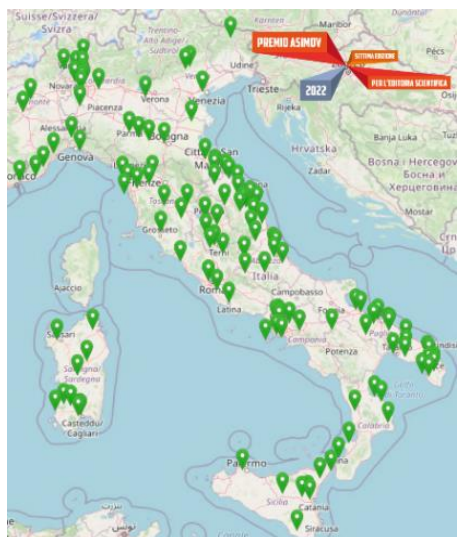

**Figure 1** – Distribution of the High Schools in Italy participating in the ASIMOV Prize.

## 3. The winners of the ASIMOV Prize

The reviews of all the students are anonymously judged by the National Committee on the basis of their originality and of the content validity, also performing checks with an anti-plagiarism software. Each Regional Committee organizes local ceremonies to award the 15 best reviews at regional level. Also the reviews with the highest mark for each High School are mentioned.

The concluding act of each edition of the ASIMOV Prize consists of a final ceremony in which the awarded students are invited to present and discuss the finalist books. Finally, the author of the book that has scored the highest overall score, as determined by the votes of the Jurors, is declared winner of the ASIMOV Prize by the spokesperson, on behalf of the National Committee and of the Jury. The titles and the authors of the winning books of the seven editions are reported in Fig. 2. Since the first edition, the final ceremony has moved between different locations (except for editions 2020 and 2021, which were held online only due to Covid-19 restrictions).

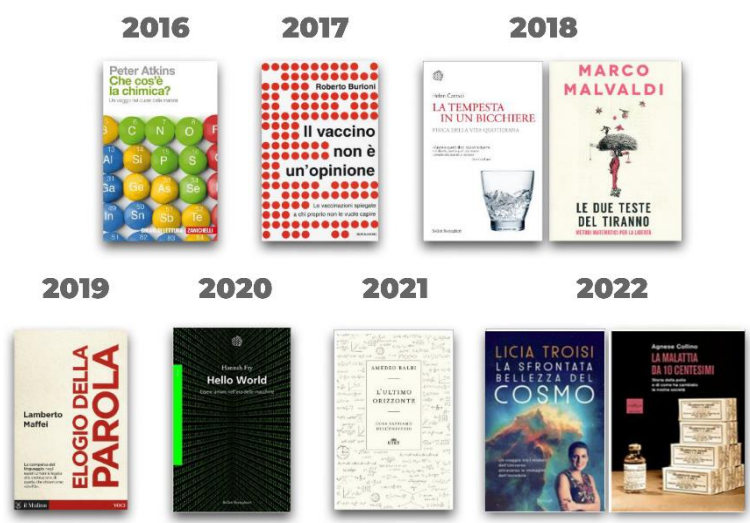

**Figure 2** – List of all the winning books in the seven editions of the ASIMOV Prize.

In years 2018 and 2022 two ex-aequo winners have been proclaimed. In particular, in 2022 the ASIMOV Prize has been assigned to two women: Licia Troisi (for "*La sfrontata bellezza del cosmo*") and Agnese Collino (for "*La malattia da 10 centesimi*"), respectively a presentation of astronomy by means of key images, and a social history of polio and how it was eradicated.

## 4. More about the ASIMOV Prize

In order to encourage a large participation in the ASIMOV Prize, various incentives have been devised for both students and teachers. In particular, students are encouraged to take part in the ASIMOV Prize and read at least one book to obtain school credits and/or recognition of young apprenticeship programs (for a total of 30 hours per book), which in Italy are presently called PCTO ("*Percorsi per le Competenze Trasversali e l'Orientamento*"), formerly ASL ("*Alternanza Scuola Lavoro*"), and aim to equip students with useful skills and competences for the purpose of employment. Awarded students win educational trips to one of the organizing bodies, as well as gadgets or bonuses for buying books, as shown in Fig. 3.

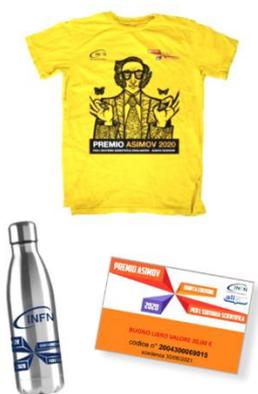

**Figure 3** – Examples of gadget for students and teachers participating to the ASIMOV Prize.

Also school teachers actively taking part to the national/regional committees acquire the recognition of successful completion of an official teachers training course (called "*SOFIA*").

The number of students participating in the ASIMOV Prize has been increasing year after year, as can be noticed from the statistics reported in Fig. 4. Moreover, among the five classes constituting the Italian High School system, the two most frequent ones are the 3$^{rd}$ and the 4$^{th}$ classes (corresponding to students aged 17-18). Participant students are equally divided between boys and girls, but the percentage of female winners is as high as 65%.

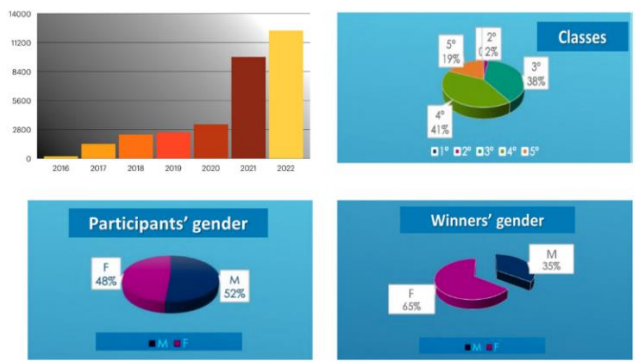

**Figure 4** – Histograms showing the growing number of students taking part in the ASIMOV Prize since 2016 to 2022 (up left), the class distribution in High Schools (up right) and the gender distributions corresponding to all the participants (down left) and to the winners (down right).

**5. Two cultures, one prize**

After seven successful editions, the ASIMOV Prize proved - once again - that the old 'two cultures' issue [3] can be effectively addressed and defused with the help of appropriate cultural and social tools. Science popularisation books were the levers we used to achieve these goals in the ASIMOV Prize, along with best practices in public involvement, starting with an active collaboration between high schools and research centres.

Indeed, from an accurate analysis of the reviews prepared for the ASIMOV Prize, it has become evident that today's students have definitively gained a new perception of scientific literature, coming into a fresh-minded contact with science and technology subjects through an innovative approach and developing a deep critical and conscious judgment. Students awarded for their reviews are encouraged to produce short videos commenting the books they have read and judged. A selection of the best videos has been published on the official YouTube channel of the ASIMOV Prize: http://youtube.com/PremioAsimov.

**6. Summary**

In seven editions since 2016, the ASIMOV Prize has been continuously growing, awarding the authors of popular science best sellers, thanks to a Jury composed of more than 12,000 students from about 270 Italian High Schools. Edition after edition, the prize has proved to be increasingly successful and is ready to reach an international relevance, as already experienced in Brazil, where an edition of the ASIMOV Prize has taken place in 2020 with encouraging results. Future plans include the possibility to replicate the initiative also in other countries.

At the time of writing, the National Committee of the ASIMOV Prize is finalizing the selection of the finalist books of the eighth Italian edition, to be held in 2023.